\documentclass[11pt]{article}  
\usepackage{amsmath,amsfonts,amssymb}
\usepackage{graphicx}
\usepackage{setspace}
\usepackage{tocloft}
\usepackage{wasysym}
\usepackage{multirow}
\usepackage{array, booktabs}
\usepackage{siunitx}
\usepackage{authblk}
\usepackage{fullpage}
\usepackage{floatrow}
\usepackage[usenames]{xcolor} 
\definecolor{xlinkcolor}{cmyk}{1,1,0,0}
\usepackage[
 colorlinks=true,    
 linkcolor=xlinkcolor,     
 citecolor=xlinkcolor,     
 filecolor=xlinkcolor,  
 urlcolor=xlinkcolor,      
 final=true
]{hyperref}
\usepackage[numbers,square,sort&compress]{natbib}

\title{High-pressure TPCs in pressurized caverns: opportunities in dark matter and neutrino physics}
\author[a,*]{Benjamin Monreal}
\affil[a]{Case Western Reserve University, Department of Physics, 10900 Euclid Ave, Cleveland, OH, USA,  44106 }

\begin{document}

\huge
\begin{raggedright}
 \textit{High-pressure TPCs in pressurized caverns: opportunities in dark matter and neutrino physics} \hfill \\[+1em]
\end{raggedright}

\normalsize

\noindent {\large \bf Author:}\\
Benjamin Monreal (Case Western Reserve U.) [benjamin.monreal@case.edu]

\newcommand{\dbd}{$0\nu\beta\beta$}
\newcommand{\hTse}{H$_2$Se}
\newcommand{\seET}{$^{82}$Se}
\newcommand{\Thtwo}{$^{208}$Tl}

\begin{abstract}
  The natural gas and hydrogen storage industries have experience creating huge, pressurized underground spaces.  The most common of these is ``solution mining'', a method for making brine-filled cavities in salt formations.   Unlike conventionally-mined underground spaces, these spaces are (a) inexpensive to construct and operate, (b) naturally serve as pressure vessels, at size scales impossible to construct in a conventional lab, and (b) permit safe use of flammable and/or toxic materials.  If various engineering challenges could be met, solution-mined caverns would allow unprecedentedly-large high pressure gas TPCs.  Lined rock caverns (LRC) may permit high pressure TPCs of considerable size in more conventional spaces.  In this whitepaper, we review some of the new physics opportunities available in these caverns and suggest an R\&D program needed to realize them.
\end{abstract}

\section{Motivations}

Why have so many underground rare-event experiments used \emph{liquid} targets rather than gas?  The seemingly-obvious answer is: ``at obtainable pressures, gases aren't dense enough to put high-mass experiments in reasonable caverns'', but this has two caveats: {\em obtainable pressures} and {\em reasonable caverns} are technological statements about conventional pressure vessels and conventional underground labs.  In this whitepaper, we explore the unconventional use of solution-mined salt caverns\cite{monreal2014underground} and lined rock caverns\cite{johanssonStorageHighlyCompressed2014} as an enabling technology for building large gas TPCs underground.   

Gases have useful microphysics properties for rare-event searches\cite{2007NIMPA.581..632N}.  Compared to cryogenic liquids:
\begin{enumerate}
\item Track lengths are resolvable even at low energies, allowing many forms of signal/background separation, including recoil vs. $\beta/\gamma$ separation for WIMP searches, $\beta$ vs $\beta\beta$ for $0\nu\beta\beta$ searches, and $e$ vs $\alpha$ for, e.g., solar neutrinos.
\item Long track lengths allow new event-direction reconstruction, allowing directional cuts on supernova, solar, or in some cases reactor and geoneutrinos.
\item Gases typically have sub-Poisson ionization fluctuations (Fano factor $F < 1$) which enables ultra-high-resolution ionization calorimetry; this is key for neutrinoless double beta decay, but also aids in the tagging and assay of radioactive backgrounds.
\end{enumerate}
Today, these microphysics advantages are notably exploited underground by NEXT\cite{NEXT:2015wlq} ($0\nu\beta\beta$ in Xe at 15--20 bar),  NEWS-G\cite{NEWS-G:2017pxg} (WIMP search in H/He/Ne at 3--10 bar), plus proposals including ND-GAr \cite{Duffy:2019egx} ($\nu$-Ar interactions in Ar at 10 bar) and kiloton-scale xenon\cite{Avasthi:2021lgy} in conventional steel and copper pressure vessels.

Unfortunately, pressure vessels are difficult to scale up, particularly in underground labs with constrained access.  A rough ideal-gas scaling law (holding shapes,  materials, and safety factors constant) is that the pressure-vessel mass is always several times the contained gas mass, independent of pressure and overall size.  The situation is worse if we need clean shielding or veto material inside the pressure vessel (for example, to shield gamma emissions from the vessel's steel) and exponentially worse if we require fiducialization.  Moreover, large volumes of high-pressure gas involve different hazards than large volumes of cryogens; some targets may be virtually impossible to engineer safely in a conventional lab.  

In this paper, we describe the petroleum industry's standard (but possibly unfamiliar to physicists) giant underground pressure vessels: the widely-commercialized solution-mined salt cavern (section \ref{sec_sm}) and the newer lined rock cavern (section \ref{sec_lrc}).   In these caverns, the {\em entire underground space}---not a pressure vessel within it---can be pressurized.  The caverns can be made and at depths, sizes, and project costs well-suited for TPCs, and at scales which permit self-shielding.  This whitepaper lays out the physics opportunities (section \ref{sec_opp}) available from giant gas TPCs, outlines some of the general engineering solutions for deploying them (section \ref{sec_eng}, and lays out specific detector ideas (section \ref{sec_exp}) in hopes that it motivates the community to work on further brainstorming, R\&D, and detailed proposals.

\section{Pressurized cavern technologies}\label{sec_cav_tech}

\subsection{Solution-mined caverns}\label{sec_sm}

\begin{figure}
\floatbox[{\capbeside\thisfloatsetup{capbesideposition={right,center},capbesidewidth=4cm}}]{figure}[\FBwidth]
{\caption{Examples of existing solution-mined salt caverns, lined rock caverns, and selected underground lab excavations (real and proposed) for scale. Two drawings of the Eminence cavern show its rapid shrinking due to salt creep when underpressurized.  Salt cavern data from~\cite{Warren2006}).}\label{fig:test}}
{\includegraphics[width=8.5cm]{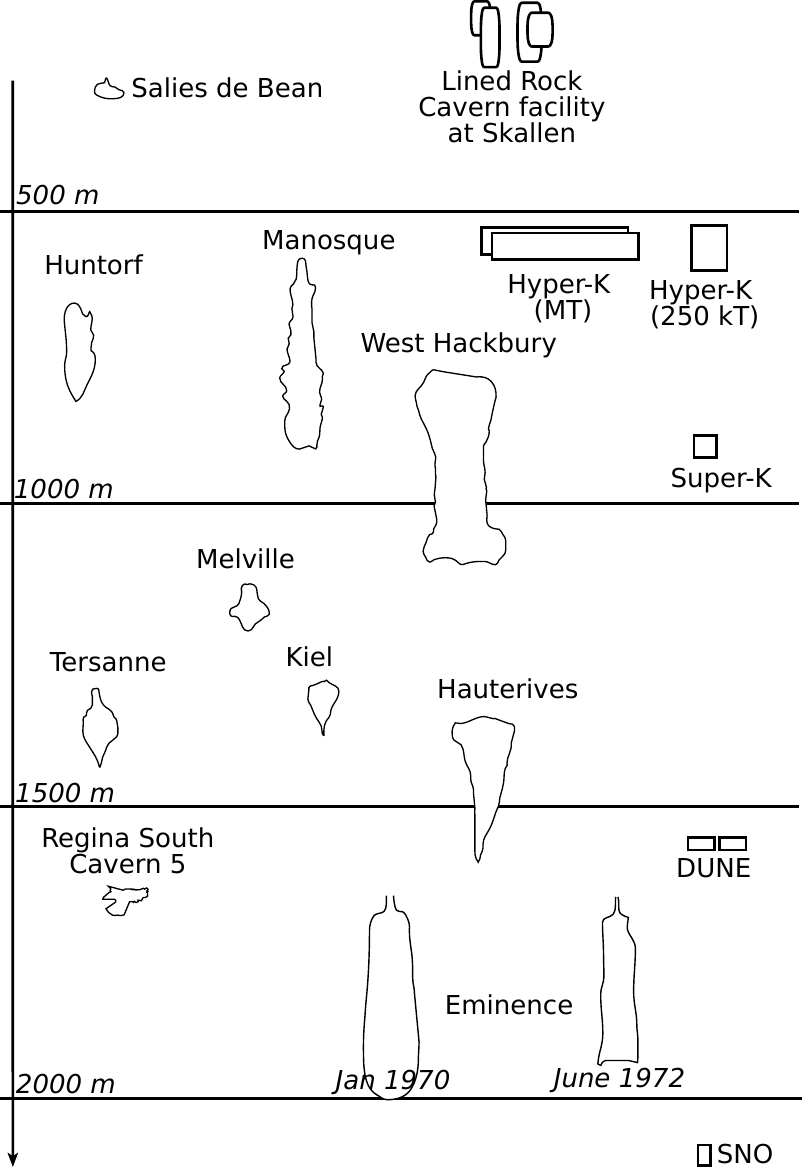}}
\end{figure}

Solution-mined caverns are created without human underground access.  A well is drilled into a large salt formation, which might be a salt dome or a salt bed\footnote{Salt domes are large pillar-like underground intrusions or diapirs of salt, while salt beds are intact sedimentary layered structures.  Beds are the more familiar formation to physicists; the Boulby, WIPP, and historic IMB underground labs were mined (mechanically, not solution-mined) in bedded salt. However, salt domes---particularly common on the US Gulf Coast, northern Europe, and the Persian Gulf---host the largest caverns.}.  Fresh water is pumped down the well to dissolve the salt; the brine is withdrawn and recycled, sold, or discarded.   Careful management of the water injection and withdrawal pipes, and use of cover gases, allow the creation of a brine-filled cavern of some desired shape and size. In typical petrochemical storage use, products (natural gas, liquid hydrocarbons, hydrogen) are pumped into the cavern, displacing the brine, aiming for a roughly-constant wellhead pressure, usually near or slightly above the hydrostatic pressure of 100 bar/km.  

The caverns may be truly immense, with the internal pressure providing some mechanical support for the walls and roof.  Caverns of 80~m diameter and 500~m high, enclosing $2.5\times 10^6$~m$^3$, are fairly standard in Gulf Coast salt domes.  In salt beds, cavern sizes are usually smaller and shapes less regular, but it still easy to obtain tens of meters of clear volume.  Operating a cavern with inadequate pressure may allow the salt to creep slowly closed\cite{coates}; this creep may be negligible for shallow (1000 m) caverns, a moderate effect (1\%/y) at 2000 m, but quite rapid (10\%/y) at 3000 m.   Thus, we should view ``medium-pressure'' experiments---say 1--50 bar---as preferring shallow caverns (say 0--1500 m).  Deeper caverns would be reserved for smaller experiments (which maybe occupy only a small fraction of a cavern and don't care if it is creeping shut) or for very high pressure (200+ bar) sufficient to support the cavern itself.  Famous examples of salt-cavern collapses causing surface subsidence, like Bayou Corne, are well studied; engineers and regulators claim to be able to avoid them in the future.  The "rockfall" risks inside working storage caverns have not been well studied.   

In standard uses, solution mined caverns are never entered by personnel or by large equipment; they are connected only to the drilled and cased well, which is typically 25--45~cm diameter but plausibly larger. Equipment, typically sonar logging equipment and fluid tubing, but in one case a remotely-operated inspection submarine\cite{ballou}, can be lowered down the well.  Is wellbore-only-access a constraint on future physics uses, too?  If so, we are looking for detector technologies which can be engineered to fit down a well---maybe not a conventional 45~cm well but something (100--200~cm, perhaps?) affordable with familiar shaft sinking methods.  If not, perhaps we can devise a mix of conventional and unconventional mining; for example, we might mine conventionally into the upper salt or caprock, and do preparatory work there before solutioning a cavern underneath it.  

\subsection{Lined rock caverns}\label{sec_lrc}
In Sweden, near Halmstad, a storage facility for natural gas uses lined rock caverns (LRC)\cite{johanssonStorageHighlyCompressed2014}. This commercial site has mined out several cylindrical caverns in rock (60--100~m high, 35-40~m diameter) and lined them with steel to store natural gas at 300~bar. The cavern was prepared by excavating, shotcreting (including drainage pipes to manage groundwater during construction), and pre-placing rebar.  Next, a steel liner was welded together from 12mm thick carbon steel plates.  Grout was injected behind the steel, forming a reinforced concrete contact surface that conformed to, but did not bond with, the steel. When the cavern is pressurized, the liner deforms outward and transfers its load to the rock via the concrete.  Due to the sliding interface, the force on the concrete is always compressive rather than shearing.  In this way, it is the compressive strength of the rock (rather than hoop stress in the steel, which would not sustain even 1 bar on its own) that withstands the pressure.   

A fully pressurizeable cavern, constructed and lined on this model, could be the construction site for a very large TPC of any desired shape and materials.  Inside the cavern, we can build a TPC inside of a thin, low-radioactivity gas envelope. Between the gas envelope and the cavern lining/wall, we can install clean water shielding tanks, veto systems, etc..  As long as we are careful to fill the TPC envelope (with, say, xenon) and backfill rest of the cavern (with, say, nitrogen) synchronously, there is no net pressure drop across the envelope; it only needs enough mechanical strength to withstand gravity and (if the fill gas has a different density than the backfill)  buoyancy.

Compared to a normal underground lab experiment, this does not obviate all pressure/instrumentation challenges; ancillary equipment in the cavern (veto system PMTs, thermal insulation for chilled SIPMs, etc.) now needs to be under high pressure.  Compared to the Halmstad LRC facility, which has only a small manway, extra attention must be paid to a pressure hatch for personnel and equipment that can mate with the liner without concentrating stresses.  Moreover, the Halmstad facility does not require underground personnel and it is not clear what hazard level would be associated with occupation of the underground tunnels while the cavern is pressurized. 

A 1997--2002 DOE-funded study\cite{osti_774913} estimated the "belowground" construction cost of a shallow (100--200~m) four-cavern facility, with 300,000~m$^3$ total volume, would be \$125M.

\section{Physics opportunities in giant gas TPCs}\label{sec_opp}

The physics opportunities, and their different approaches to targets and to salt-cavern/LRC constraints, may be roughly grouped as follows: 

\begin{enumerate}
\item Targets which would be useful at ton-scale, but whose hazardous properties make them challenging to use in (in any phase) a conventional pressure vessel parked in an existing cavern. These include flammable/explosive H$_2$ (for light WIMPs) or toxic gases including H$_2$Se, TeF$_6$, GeH$_3$, MoF$_6$ (for \dbd). 
\item Targets which a switch from liquid-phase to gas-phase is well motivated by physics but stymied by conventional pressure-vessel constraints: Xe (particularly \dbd~but potentially WIMPs), He and Ne (WIMPs, solar neutrinos), possibly Ar (WIMPs, some long-baseline phenomena).   
\item Targets where, independent of the phase, the interesting cases are 50-kiloton- to megaton-scale but the large-scale cryogenics and/or the enormous excavated space requirements are prohibitively expensive.  If an attractive flagship-scale project of this sort can be defined, the smaller projects motivated above could be seen as technology pathfinders.  Targets might include H$_2$ and CH$_4$ (reactor/geoneutrinos, long-baseline); CO$_2$, Ar (proton decay, long-baseline).
\end{enumerate}

\section{Engineering solutions for down-well deployable TPCs}\label{sec_eng}

Today's underground experiments are extremely precise, handcrafted objects with thousands of carefully-positioned parts.  Conventional TPC assemblies can be used in a lined rock cavern.  To use a solution-mined cavern, we have to stretch our imaginations.  Is it realistic to imagine building a detector that can be squeezed down a narrow well, then deploy and operate without any human hands-on access?  How?  There are some precedents from familiar experiments that may serve as a guide, though obviously a large and novel engineering effort is needed.

\begin{enumerate}
\item {\bf Inflatable TPCs} Spherical and cylindrical TPCs are probably the simplest to adapt to down-well deployment.  Here, all of the sensitive/fragile instrumentation components are on a small anode assembly; the cathode is uninstrumented.   The obvious solution is for the cathode to be a balloon; it can be furled up to deploy down the well into the cavern, then inflated with the target gas; the anode array would be designed to go down the well in one piece.   Spherical TPCs are discussed in \cite{Andriamonje:2009zz} and implemented in \cite{Bouet:2020lbp, Arnaud:2017bjh}; cylindrically-symmetric, radial-drift TPCs should have many of the same properties.
\item {\bf Deployable} Conventional TPCs use linear drift between large, flat electrodes; such a structure doesn't fit down a well.   Many spacecraft have a similar problem in which a desired extended structure (an antenna, solar panel, sunshield, or radiator) doesn't fit in a spacecraft fairing.   In these cases, the flat structures are designed to expand (or, rarely, inflate) from a hinged, rolled, or furled configuration.   Via similar engineering, rolled or folded electrode arrays could be expanded into a TPC inside a gas balloon in a cavern.
\item {\bf Cavern-filling subunits} The SNO experiment, in its third phase, deployed an array of neutron-capturing $^3$He proportional counters (NCDs) by dropping them down the neck of the ultrapure acrylic water vessel.   A remotely-operated submersible picked up each NCD and delivered it to an anchor point in the water.  Relatedly, KM3NeT uses industry-standard remotely operated submersibles (ROVs) and equipment to install cabled PMT strings in the deep Mediterranean.   A large salt cavern experiment might be built similarly.  Self-contained detector subunits---possibly small inflatable TPCs---would be designed to fit down the well one by one and be (at least during the assembly process) neutrally buoyant in the cavern fluid.  An in-cavern ROV would roughly position, secure, and cable the subunits.
\item  {\bf In-cavern detector racking}  Though DUNE is large and complicated, its main active components are tall, narrow anode plane arrays (APAs).  DUNE's assembly process brings fully-assembled, pre-tested APAs into the cryostat through a narrow port, then rolls them into place on preinstalled rails.   One can imagine a related assembly sequence in a cavern: first, with the cavern flooded with water, an ROV assembles a rail system out of neutrally-buoyant components.  The ROV is withdrawn, the cavern is dewatered and dried. Then, tall and narrow APA assemblies are lowered down the dry well into the cavern one by one, where they mate with the racking system.
\end{enumerate}

\section{Candidate Experiments}\label{sec_exp}

\subsection{\texorpdfstring{\dbd~in \hTse}{0vbb in H2Se}~and related gases at 1~t scale and beyond}

\seET~is a double beta decay isotope---one of the few with an endpoint energy above \Thtwo~backgrounds---which has been measured only at the kg scale by NEMO-3 and CUPID-0; CUPID may provide a future route to ton-scale but not beyond.  However, Se exists in a convenient form, \hTse, which we speculate could be used in a TPC, either as a gas (up to 9 bar at room temperature) or a liquid.  By analogy with H$_2$O we expect that it requires negative-ion drift, but detailed measurements or molecular calculations are lacking; however, the physics opportunities are attractive enough that we will speculate that it works.   

\hTse, like the other candidate gases we discuss below, is toxic and flammable; this author finds the idea of bringing tons of it into a multipurpose underground lab daunting, although we welcome suggestions on how to do so.  A lined rock cavern attached to a general-purpose underground lab {\em may or may not} be able to establish the desired safety level.  A solution-mined cavern site would not have this problem at all; since only the well-ventilated surface labs are occupied.

Against external backgrounds and solar neutrinos, we anticipate extremely powerful track-topology background rejection---better than obtainable in xenon---due to (a) reduced large-angle electron scattering in this lower-Z medium and (b) the likely "diffusionless" spatial resolution of a negative-ion TPC.  High energy resolution might be obtained with the single-primary-counting feature of very slow drift, as in a time expansion chamber.  

An idealized zero-background 3 ton-year measurement of enriched \seET~could reach $\mathcal{O}(5)$ meV neutrino mass sensitivity.  At 9~bar, a single-phase gas \hTse~TPC would have a 40 kg/m$^3$ density and require a 31 m$^3$ fiducial volume.  It is not, unfortunately, easy to divide this into wellbore-sized mini-TPCs, since typical electron tracks span several tens of centimeters and we wish to fully contain them.  Instead, we envision a system that "folds" (with, we hope, only moderate mechanical complexity) into a wellbore-compatible compact state and "unfolds" inside a gas-filled balloon once in the cavern.

Note that 9~bar is a safe operating pressure in a shallower cavern but not a deeper one; to go deeper, we might opt for 9 bar \hTse~partial pressure in a mixed \hTse /He TPC; or abandon the single-phase gas approach and use a compact two-phase TPC, albeit anticipating a loss of calorimetric resolution and therefore some 2$\nu\beta\beta$ backgrounds. 

Although \seET~seems the most promising, we can also consider other acutely-toxic, flammable, corrosive gas and liquid forms of the \dbd~candidates in Table \ref{dbd_table}.  None of these are familiar detector gases and their drift properties, electron attachment/detachment coefficients, gain stability, etc., are not well known.  (Dimethyl cadmium, though interesting from a physics perspective, we omit as too hazardous even for small-scale R\&D.)  We add Xe for comparison; note that Xe although has a ton-scale path forward in conventional pressure vessels, pressurized caverns permit this program to expand to the limit allowed by xenon supply (possibly kiloton-scale\cite{Avasthi:2021lgy}.

\begin{table}
\begin{tabular}{c|c|c|c|c|c}
  Target  & Endpoint & working point   &\multicolumn{2}{c|}{m$_\nu$ = 50 meV} & m$_\nu$ giving \\ 
           &  keV     & P, T, $\rho$        &  Events/t/y & Events/m$^3$/y & 1 event/t/y \\ 
\hline 
H$_2$Se & 2998 & 9 bar, 273 K, 33.7 kg/m$^3$ & 22 & 0.7 & 10 meV \\ 
MoF$_6$ & 3035 & 1 bar, 320 K, 9.4 kg/m$^3$ & 26 & 0.2 & 9 meV \\ 
TeF$_6$ & 2527 & 10 bar, 273 K, 107.5 kg/m$^3$ & 3 & 0.4 & 27 meV \\ 
GeH$_4$ & 2039 & 50 bar, 273 K, 178.5 kg/m$^3$ & 1 & 0.3 & 36 meV \\ 
Xe & 2459 & 50 bar, 273 K, 303.4 kg/m$^3$ & 8 & 2.5 & 17 meV \\
\end{tabular}
\caption{Five candidate gases for \dbd~search TPCs; the correspondence between event rates and neutrino masses is given for roughly-mid-range matrix elements.} \label{dbd_table}
\end{table}

\subsection{Solar neutrinos and dark matter in Ne at 100 T scale}

\begin{figure}
\floatbox[{\capbeside\thisfloatsetup{capbesideposition={right,center},capbesidewidth=5cm}}]{figure}[\FBwidth]
{\caption{Conceptual arrangement of solution-mine-deployed, gas-shielded neon TPC. 10~m of denser gas, probably argon, shields external gamma rays from U/Th/K in the salt.  A tough nylon balloon holds back the buoyancy of a less-dense, cleaner Ar:CH$_4$ shield, which may be instrumented as a veto.  The neon active volume is contained by a thin cathode balloon, with the instrumented anode suspended in its center.  In-situ high voltage generation allows multi-megavolt drift.}\label{fig:ne_det}}
{\includegraphics[width=9cm]{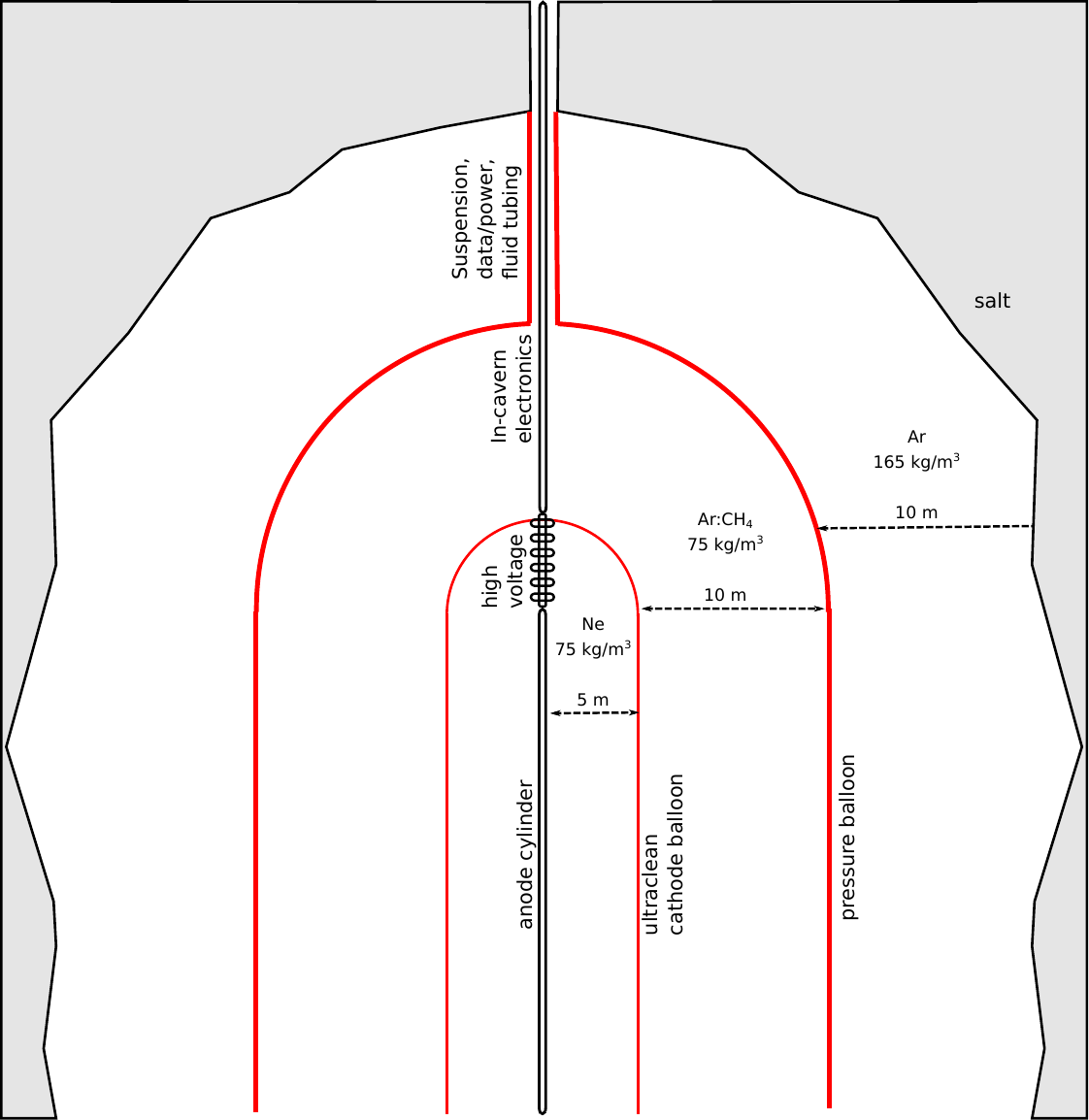}}
\end{figure}

The solar neutrino spectrum was first identified in low-energy-threshhold detectors (Homestake, GALLEX) but most conclusively studied at higher energy threshholds (SNO, SuperK).  In liquid scintillators, the lower-energy solar neutrino spectrum is accessible (including $pp$ and CNO neutrinos), but runs into irreducible or nearly-backgrounds including $^{14}$C and cosmogenic $^{11}$C, and, despite heroic radiopurification, $^{210}$Bi.  Will we ever do solar neutrino spectroscopy with lower backgrounds than Borexino?  Equally importantly, can we do so on a modest budget?  Consider a neon gas TPC\cite{monreal2020poster} as an alternative.  In a TPC, unlike a scintillator, electron, positron, and alpha events can be tagged by the track length and topology.  Electron recoil direction measurement \cite{Bonventre:2018hyd} separates solar neutrinos from isotropic backgrounds, significantly relaxing the background demands.  

A radial-drift cylindrical gas TPC seems to have the properties needed for this measurement.  The large surface area and lack of self-shielding admits more external gamma backgrounds than a conventional liquid-phase experiment, but the low-density, low-Z medium allows unprecedented tracking which we believe may compensate.  

Consider a TPC 80~m tall, with a 50~cm rigid anode array hanging coaxially in a 10~m diameter cathode balloon, accommodating 500~T Ne at 100~bar.  In a solution-mined cavern, the balloon is surrounded by a wide (10--20 m) gas-filled space between the detector and the cavern wall.  This space might be open and carry a single neon-density-matched shielding gas, or might have additional balloons and a nested structure allowing denser gases. In a lined rock cavern with higher excavation costs, it might be more cost-effective to shield with water or paraffin.

With the anode at 250kV, electrons can drift the full radial distance in under 5~ms.  (This can be cut in half by increasing the anode voltage to 1 MV, or by switching to a faster gas like Ne:CH$_4$ 97:3) a track born at the cathode will suffer about 5~mm of transverse and longitudinal diffusion---enough to permit diffusion measurements to provide r-coordinate information but still allowing excellent track reconstruction for MeV electrons.  The spatial resolution provides very good background rejection, in addition to the sun-direction cut.  In the low-Z medium of neon, external gammas nearly always (by a factor of 10$^4$) produce multisite Compton events rather than single-site photoelectric electrons.  There are no $\beta^-$-emitting spallation products lasting longer than 10~s, allowing a reasonable muon-follower veto; longer-lived $\beta^+$-emitters will have a $\beta^+$ annihilation signature.  The impact of radon requires further study; can $\beta^-$-emitting radon daughters float around in the gas?  

The same apparatus can do a WIMP search, although a better geometry might be, say, $\diameter 10 \times 25$~m at 300~bar.  Pointlike, few-keV ionization events would provide the WIMP signature, without LXe-quality recoil/beta discrimination but with much lower quenching. For high-mass WIMPs, scaling by a coherent-scattering factor of $20^2/130^2$ suggests 500~T Ne has a similar event rate to 10~T Xe.  For lower-mass WIMPs, scaling from proposed 25~kg target of \cite{Lippincott:2017yst} suggests we might do neutrino-floor-limited WIMP searches down to 2--3 GeV.  

\subsection{Geoneutrinos and antineutrinos in \texorpdfstring{H$_2$}{H2} at 3~kT scale}
Hydrogen gas would be a "golden" target for many physics studies if it could be wrangled into a large detector.  In inverse beta decay $\bar{\nu_e} + \mathrm{p} \rightarrow \mathrm{n} + \mathrm{e}^+$ in scintillators, in conventional detectors all the physics information comes from the $\beta^+$; the neutron, captured later, serves only as a delayed tag.  However, with better granularity we can exploit two facts: (a) that the neutron picks up most of the neutrino's momentum (and typically 10 keV kinetic energy) and can be used to tag direction, and (b) although the positron's kinetic energy carries all the of the neutrino calorimetric energy, and the positron annihilation, though prompt, leaves spatially-displaced energy deposits that distinguish $\beta^+$ from $\beta^-$.  

In an ultra-high-spatial resolution detector, an inverse beta decay event has a wealth of information: a positron track; two positron-annihilation gamma conversions; a series of neutron-thermalization hits, and finally a (displaced and delayed) neutron capture.  We expect that it is very, very difficult for any ordinary radioactive background to mimic the full signature.  For example, cosmogenic $^9$Li $\rightarrow e^- + \bar{\nu} + n + 2\alpha$---a genuinely irreducible background in a large scintillation detector---lacks the IBD positron annihilation signature.  

A low-threshold detector could exploit the antineutrino direction tag in interesting ways: 
\begin{enumerate}
\item Separation of crust from mantle geoneutrinos; separation of reactor backgrounds from geoneutrinos
\item Isolation of specific reactors in a multi-reactor signal
\item Tight directional clustering of supernova antineutrinos
\end{enumerate}

To start with, let us focus on H$_2$.  Its advantages are that it is almost perfectly free of long-lived cosmogenics, it simplifies the detection of neutron thermalization; its low density makes $\beta$ tracks typically quite long and resolvable.  The downside is a lack of self-shielding and imperfect event containment.  We can add stopping power, at the cost of maybe reintroducing cosmogenics, by blending in CH$_4$ or neon.  Doping with $^3$He would sharpen the neutron tag, but it is unlikely to be available in the quantities needed. Doping with $^{10}$B (as diborane or BF$_3$) is worth studying as an alternative.

A 3~kT H$_2$ target has the same hydrogen content as THEIA's proposed 25 kT H$_2$O; this is the scale where we expect $\mathcal{O}(1)$ IBD event from a supernova in Andromeda, or $\mathcal{O}(200/y)$ mantle geoneutrinos.  In solution mining, 3 kT is sane from cavern-engineering and even hydrogen-storage standards; an active ConocoPhilips hydrogen-storage cavern in Clemens Dome, Texas has a volume of 580,000 m$^3$ (50~m diameter, 300~m tall) and can store 5.5~kT H$_2$ at 135~bar.  Could we turn a similar cavern into a detector?  First, we note that hydrogen is a "slow" drift gas and we don't know its purity limitations.  We can contemplate scaling up the single-anode cylindrical neon detector above, going to $40$~m diameter and 270~m height.  To keep drift times short, the central anode needs to be bigger, and therefore needs to be a multi-piece or expanding assembly, possibly 4~m diameter at 2 MV, or a 2~m at 5 MV.  To work with shorter drift times we would probably switch to a multi-anode cavern.

Scaling up further, larger caverns ($10^6$ m$^3$) and higher pressures (300~bar) might allow targets up to 20~kT H$_2$ (equivalent to 180 kT H$_2$O). Scaling down, it would be interesting to study the physics case for smaller, kiloton-H$_2$O equivalent H$_2$ projects for nuclear security.

Worldwide, nuclear reactors sited over or near salt beds are fairly common, currently including Ohio, Michigan, the US Gulf Coast; Hartlepool in the UK; northern Germany, southeastern France, Catalonia, eastern Ukraine, scattered sites in Russia and China; Bushehr in Iran; and the future Akkuyu plant in Turkey. 

\subsection{The largest caverns and the cheapest gases: proton decay at megaton scale}

Searches for proton decay continue to be interesting and powerful.   SuperKamiokande data limits the lifetime to $t > 1.6\times10^{34}$ y for $\mathrm{p} \rightarrow \mathrm e^+ \pi^0$ (one event per 50 kT/y) or, for a promising but less-WC-friendly mode, $t > 10^{33}$~y  for $\mathrm{p} \rightarrow \mathrm e^+ K^0$ (one event per 3 kT-y).  Further progress requires detectors with quite large masses, but would also benefit from TPC-like rather than WC-like reconstruction; note that those SK limits are obtained after 450 kT-y of exposure.   

The most common large solution-mined caverns in the Strategic Petroleum Reserve have a volume of $1.8\times10^6$~m$^3$, accommodating perhaps 250--500~kT of dense, inexpensive gases like CH$_4$, CO$_2$, or (not really ``inexpensive'' at this scale) Ar; at these pressures liquid CO2 is also possible.  Is it possible to instrument this much gas?  It is not clear, but the huge payoff---HyperK-like or larger exposure with TPC-like tracking---seems to merit further study, as do non-proton-decay physics goals like atmospheric or long-baseline neutrino physics.  Recall that the caverns themselves are inexpensive (possibly \$15--30M, i.e. $\mathcal{O}(5)\%$ of the cost of a compararable rock cavern), as are the gases (\$50M plus purification for CH$_4$; more for argon) leaving budget flexibility to focus on engineering challenges.  

For one approach, consider an inflatable tube-shaped detector unit 0--50~m long, 2~m diameter and holding 20--50~T of gas.  cylindrical surface serves as a cathode, while a membrane stretched across a diameter carries a flat sensitive anode (perhaps microstrip-like or RPC-like), field-shaping electrode strips, and readout electronics.  The balloons lay on their sides, and 10,000 such balloons stack in a hexagonal grid to fill the volume.  Each detector gets a custom gas fill to make it neutrally buoyant in the ambient cavern environment.  This is far from a fully-thought-out design, but is meant to illustrate that unusual geometries and novel problem-solving which the community could dive into in order to develop and realize such an experiment.  

\subsection{New far detectors in the LBNF neutrino beam}

\subsubsection{Solution-mined caverns in the Pine Salt}
The Williston Basin is a well-studied oil-bearing geological structure underlying parts of North Dakota, South Dakota, Montana, and Canada.  Its most notable thick salt bed called the "Pine Salt"; North Dakota is studying its use for petrochemical and hydrogen storage caverns.  The southern edge of the Pine Salt reaches into South Dakota, within reach of off-axis neutrinos from LBNF.  In particular, a 100~m thick bed can be found 100~km north of Lead, SD (4.4$^\circ$, substantially but not fatally reducing the neutrino flux) thinning to 30~m at a distance of 60~km (2.6$^\circ$, within the off-axis kinematic range to be studied by DUNE-PRISM) and at a depth of 2400~m.  Working pressures are likely to range from 250--500 atm.  Could a detector in the southern Pine Salt serve as an off-axis long baseline neutrino experiment?  

Consider a 30~m diameter, 30~m high cylindrical cavern at the 2.6$^\circ$ site.  This might hold 5--10 kT of high-pressure argon, yielding high event rates but with an off-axis energy spectrum.  Note that, with the intended low-cost excavation, the lower instrumentation density, and simplified infrastructure, we might specify {\em multiple caverns} to get higher target masses.  It would be interesting to study whether an off-axis argon target could complement the DUNE long-baseline physics program by exploiting the different spectrum.

If DUNE is on track to learn everything it's possible to learn in argon, could an off-axis detector provide alternative targets?  Pure H$_2$ is attractive in principle since it brings the interaction cross section nearly to zero; however, it has problems. A H$_2$ target reaches only 0.5~kT per 30~m cavern, which can of course be boosted to 10~kT scale by replication. A bigger challenge is that, due to the low density even at 500~bar, few events are fully contained; the radiation length is 20~m, and a 1~GeV muon range has a 50~m range. (While a pure-H$_2$ active target {\em backed by a calorimeter} is an attractive idea, technologically it is so different from the other proposals here that we will not try to engineer it.). For a hydrogen-rich target, 500~bar CH$_4$ is more attractive, providing a 4~kT supercritical gas target of which 1~kT is hydrogen.  Liquid-phase targets (10--15 kT) might include CO$_2$, NH$_3$, and various liquid hydrocarbons (\emph{including scintillators}) which have demonstrated long drift lengths at room temperature\cite{Dawson_2014}.  
\subsubsection{Lined rock caverns at SURF} 
If a safe way is found to construct it, a lined rock cavern at SURF might be considered as a new approach to a module-of-opportunity.  Argon at 200--300 bar has a density of 300--500 kg/m$^3$ (20\%--35\% liquid) so, at least as a first approximation, we might instrument this module with nearly-unmodified electronic, mechanical, and photosystem components from the DUNE liquid argon modules.  Unless safety considerations demand a dedicated ventilation shaft, the cost of creating a lined cavern might easily be lower than the cost of a conventional cavern and cryostat.

\subsection{Cerenkov and scintillator detectors}

This whitepaper is mostly intended to highlight giant TPCs.  However, salt caverns can also host giant PMT-based detectors---water Cerenkov, water based liquid scintillator, etc. These detectors are already {\em possible} in conventional caverns, even up to HyperK scale; are they worth attempting to adapt to the larger solution-mined caverns (using IceCube/Km3Net-like PMT strings) or lined rock caverns (using IceCube-like pressure housings and otherwise-conventional construction)?  There might be three reasons to explore these options:

\begin{enumerate}
\item {\bf Cost} Excavation is a huge cost driver item for THEIA or HyperK-scale projects.  It may be possible to simply save taxpayer money by moving a standard detector from an excavated cavern to a salt cavern, as long as the instrumentation costs do not blow up in exchange.  
\item {\bf Fluid selection} Large Cherenkov and scintilltor detectors at standard temperature and pressure are limited to water and hydrocarbon targets.  In a high-pressure environment, CO$_2$, N$_2$, Ar, or more exotically Cl$_2$ or SO$_2$ might emerge serve as Cerenkov and/or scintillation media. A unique aspect of gas/supercritical media is the Cherenkov threshold and angle can be tuned by pressure adjustment.
\item {\bf Site flexibility} Large experiment planning is often tied to a narrow list of sites tied to existing mine infrastructure.  Salt beds are widespread and spatially extended.  A detector that needs multiple sites (a geoneutrino transect wishing to sample different points on the crust? An decay-at-rest experiment needing multiple baselines but unable to move the source?) might exploit the flexibility of solution-mine project siting.  
\end{enumerate}

\section{R\&D needs for a path forward}

In preliminary work on this topic, we have focused on the big picture: what gases might do what physics at what scales.  At this stage do not have a full physics-sensitivity study of any one detector.  Any or all of the basic sketches shown here may have showstopper detector-physics problems.  The mechanical sketches, particularly for down-well deployable systems, might prove to be unbuildable, or unacceptably costly, in some way.  Some of the proposed cavern uses may fail on some unforseen point of salt mechanics, well management, or drilling.   

For this to move forward, we believe four things need to happen, roughly in parallel and in communication:

\begin{enumerate}
\item From the physics community, we need more people to brainstorm about desirable detector configurations, identify the relevant detector-physics uncertainties, and devise an R\&D program to reduce these uncertainties.  Early physics-oriented work might include ultra-high-pressure gas gain measurements, basic drift measurements of \hTse, GEANT4 modeling, etc.
\item We need mechanical-engineering design work to define some of the size and space constraints.  For the down-well deployments, we feel that the most pressing mechanical question is to add realistic materials and a deployment sequence to the nested-balloon proposal, and spec out the options (and risks) for delivering multicomponent anodes.  For lined rock caverns, it's important to establish whether the caverns can coexist safely with multipurpose labs or whether they need to be operated from the surface. 
\item From the funding agencies: this whitepaper's basic concept is cross-disciplinary, spanning several frontiers of astroparticle physics.  Unlike the 2001-2003 Homestake proposal process, it does not even propose a particular site.  Thus, at least at an initial scale, the agencies might see
\begin{itemize}
\item Requests for detector-R\&D funding for technologies whose underground home doesn't exist yet
\item Requests for engineering-R\&D funding to solve underground siting/mining problems for which no particular physics goal has been specified.
\end{itemize}
To realize this program, in particularly its promise of a future underground program at low taxpayer costs, we request that the agencies fund proposals in some of these development areas despite existential R\&D risk in others.
\item Physicists and engineers should begin design consultation with drillers, salt cavern engineers, and cavern owners, a process which has already somewhat started\cite{monreal2016smri}.   We note that most expertise here resides in oil-industry consultants rather than in academia or (with the exception of Sandia) national labs.  While many early-stage detector proposals make a lot of progress {\emph prior to any grant funding for proposal development}, due to the need for this specialized engineering support, that may be not be the case here.  We would like to encourage relationship-building between the DOE Office of Science, Office of Fossil Energy, and EERE on behalf of these projects.
\end{enumerate}

Since salt caverns and salt formations are already common, there is a \emph{very} low-risk path forward starting with small detectors.  Small detector prototypes and/or engineering test articles can be lowered into any of hundreds of existing, disused, flanged-off salt caverns in the US more or less immediately; as long as (a) they fit down the typically-smaller wellbores (8'', say) and (b) the device can operate in the ambient cavern brine and does not require a complex gas fill or pressure excursion.   A DOE-supported test site, ideally with an onsite wireline rig, would facilitate this.   It is possible that existing caverns and wells can support some new-physics-capable experiments.  A small interdisciplinary group at Case Western Reserve University has been discussing the possibility of a research borehole on or near campus.

\bibliographystyle{hunsrtnat}
\bibliography{references}
\end{document}